\documentclass[prb,twocolumn,aps,showpacs,floatfix]{revtex4}
\usepackage{amsmath}
\usepackage{graphicx}
\usepackage{subfigure}

\begin{document} \bibliographystyle{apsrev}

\title{Effect of electron-lattice interaction on the phase
separation in strongly correlated electron systems with two types
of charge carriers}

\author{A.O.~Sboychakov} \affiliation{Institute for Theoretical
and Applied Electrodynamics, Russian Academy of Sciences,
Izhorskaya Str. 13, Moscow, 125412 Russia}
\author{A.L.~Rakhmanov}
\affiliation{Institute for Theoretical and Applied
Electrodynamics, Russian Academy of Sciences, Izhorskaya Str. 13,
Moscow, 125412 Russia}
\author{K.I.~Kugel} \affiliation{Institute
for Theoretical and Applied Electrodynamics, Russian Academy of
Sciences, Izhorskaya Str. 13, Moscow, 125412 Russia}

\begin{abstract}
The effect of electron-lattice interaction is studied for a strongly correlated electron system described by the two-band Hubbard model. A two-fold effect of electron-lattice interaction is taken into account: in non-diagonal terms, it changes the effective bandwidth, whereas in diagonal terms, it shifts the positions of the bands and the chemical potential. It is shown that this interaction significantly affects the doping range corresponding to the electronic phase separation and can even lead to a jump-like transition between states with different values of strains.
\end{abstract}


\date{\today}

\maketitle

\section{Introduction}

A typical feature of the strongly correlated electron systems is the formation of inhomogeneous states~\cite{dagbook,dagsci}. The nature of such inhomogeneities based on the electron correlations. However, their specific manifestations can include the effects of different degrees of freedom existing in the solids: spin, charge, orbital, and lattice~\cite{khom}. A strong electron-lattice coupling plays a fundamental role in such actively studied systems as high-temperature superconductors, manganites, cobaltites, and other related materials~\cite{egami1,egami2}. An important characteristics of all these materials is a complicated electronic structure involving two or more conduction bands. The simplest model allowing appropriate description of electron correlation effects is Hubbard model and its multiband generalization~\cite{pwand,varma,wagner}.

In Ref.~\onlinecite{tb1} it was demonstrated that existence of two bands in the Hubbard model gives rise to the existence of the electronic phase separation even in the absence of any type of additional factors (spin, charge, orbital, etc.). However, these factors are necessary to give a realistic description of the particular physical systems. For example, taking into account spin and orbital variables allows a detailed picture of the phase diagram for manganites~\cite{tb2,tb3}. Including in the model the possibility of the spin-state transitions gives an explanation of the phase separation in cobaltites~\cite{ooco,ishih}. An important and vast field of research is a problem of existence of
inhomogeneities in high-temperature superconductors, especially in cuprates~\cite{egami2}. In this field, the multiband Hubbard model provides some insight in the properties of the cuprate superconductors.

Electron-lattice coupling was incorporated in the multiband Hubbard model to describe electronic structure of cuprates and manganites~\cite{lor,poliak,suph,indusi}.
In these papers, the main emphasis was put on the polaron effects or the influence of the electron-phonon interaction on electron pairing. Here we use a similar type of the electron-lattice coupling to analyze its effect on the formation of the inhomogeneous states within two-band Hubbard model.

\section{Model and a qualitative analysis}

Let us consider a strongly correlated electron system with two
types of charge carriers, $a$ and $b$, interacting with lattice
strains $u$ (static dispersionless phonons). Here, we limit
ourselves to the case of small strains, $|u|\ll 1$, when the
theory of elasticity is applicable.

The Hamiltonian of such a system can be written as
\begin{equation} \label{ham}
H=H_{el}+H_U+H_{el-ph}+H_{ph}.
\end{equation}
Here, $H_{el}$ corresponds to the energy of charge
carries  without taking into account the interaction between them
\begin{equation}\label{kin}
H_{el}=-\sum_{\langle\mathbf{i}\mathbf{j}\rangle,\sigma}
\left(t^{a}a^\dag_{\mathbf{i}\sigma}a_{\mathbf{j}\sigma}
+t^bb^\dag_{\mathbf{i}\sigma}b_{\mathbf{j}\sigma}\right)- \Delta
E\sum_{\mathbf{i}\sigma}n_{\mathbf{i},\sigma}^b, \end{equation}
where $a^\dag_{\mathbf{i}\sigma}$ and $a_{\mathbf{i}\sigma}$
($b^\dag_{\mathbf{i}\sigma}$ and $b_{\mathbf{i}\sigma}$) are the
creation and annihilation operators for $a$ ($b$) electrons at
site $\mathbf{i}$ with spin projection $\sigma$,
$\langle...\rangle$ means the summation over the nearest
neighbors, $t^{a}$ and $t^{b}$ are the corresponding hopping
integrals, $\Delta E$ is the energy shift between $a$ and $b$
bands, and $n^a_{\mathbf{i}\sigma}=a^\dag_{\mathbf{i}\sigma}
a_{\mathbf{i}\sigma}$ and $n^b_{\mathbf{i}\sigma}=b^\dag_
{\mathbf{i}\sigma} b_{\mathbf{i}\sigma}$ are the number operators
for $a$ and $b$ electrons, respectively.

The $H_U$ term corresponding to the on-site Coulomb repulsion of
charge carriers has the form \begin{equation}\label{Q}
H_U\!=\!\frac{1}{2}\!\sum_{\mathbf{i},\sigma}
\left(U_an^a_{\mathbf{i}\sigma}
n^a_{\mathbf{i}\bar{\sigma}}\!+\!U_bn^b_{\mathbf{i}\sigma}
n^b_{\mathbf{i}\bar{\sigma}}\right)\!+\!\frac{U_{ab}}{2}\!
\sum_{\mathbf{i},\sigma,\sigma'}n^a_{\mathbf{i}\sigma}
n^b_{\mathbf{i}\sigma'}, \end{equation} where $U_a$, $U_a$,  and
$U_{ab}$ are the energies of Coulomb repulsion between two $a$,
two $b$, and one $a$ and one $b$ electrons at one lattice site,
respectively, and $\bar{\sigma}$ means the spin projection with
the sign opposite to that of $\sigma$. We assume that the on-site
Coulomb repulsion is large, $U_a,U_b,U_{ab}\gg zt$ ($z$ is the
number of nearest-neighbor sites, $z=6$ for the simple cubic
lattice considered in this paper) and $U_a\sim U_b\sim U_{ab}\sim
U$.

The electron-lattice interaction can be chosen in the following
form
\begin{eqnarray}\label{e-p} H_{el-ph}=\sum_{\langle
\mathbf{i}\mathbf{j}\rangle,\sigma}\left(\lambda\,
a^\dag_{\mathbf{i}\sigma}a_{\mathbf{j}\sigma}u_\mathbf{j}+\lambda'
b^\dag_{\mathbf{i}\sigma}b_{\mathbf{j}\sigma}u_\mathbf{j} \right)\\
+\sum_{\mathbf{i},\sigma}\left(\lambda_a
u_{\mathbf{i}}n^a_{\mathbf{i}\sigma}+
\lambda_bu_{\mathbf{i}}n^b_{\mathbf{i}\sigma}\right),
\end{eqnarray} where $\lambda$, $\lambda'$ and $\lambda_{a,b}$ are
corresponding constants of the electron-phonon interaction and
$u_\mathbf{i}$ are the local distortions corresponding to site
$\mathbf{i}$. Thus, we take into account the effect of lattice
strains both on the on-site electron energy and intersite charge transfer.

We approximate the phonon self-energy term $H_{ph}$ as an elastic
energy of the system depending on distortions at different sites
\begin{equation}\label{el} {\cal{F}}_{\rm
elast}=\frac{K}{2}\sum_{\mathbf{i}}u^2_{\mathbf{i}},
\end{equation}
where $K$ is the elastic modulus.

We analyze the problem in adiabatic approximation considering
phonons as a classical static elastic field. To find a
self-consistent solution to the problem, we first perform
averaging of the Hamiltonian over the electronic degrees of
freedom. From the condition of the energy minimum with respect
to strains, $\partial{\langle
H(u_\mathbf{i})\rangle}/\partial
u_\mathbf{i}=0$, we obtain \begin{equation}\label{ui}
\bar{u}_\mathbf{i}=-\frac{\lambda z\langle
a^\dag_{\mathbf{i}\sigma}a_{\mathbf{j}\sigma}\rangle+\lambda'z\langle
b^\dag_{\mathbf{i}\sigma}b_{\mathbf{j}\sigma}\rangle+\lambda_a\langle
n_\mathbf{i}^a\rangle+\lambda_b\langle n_\mathbf{i}^b\rangle}{K},
\end{equation}
where $\mathbf{i}$ and $\mathbf{j}$ are the
nearest-neighbor sites.

Using Eq. \eqref{ui}, we can present the effective electron
Hamiltonian as
\begin{eqnarray}\label{Heff} \nonumber H_{\rm
eff}&=&-\sum_{\langle \mathbf{i}\mathbf{j}\rangle,\sigma}
\left[\left(t^a-\lambda
\bar{u}_\mathbf{i}\right)a^\dag_{\mathbf{i}\sigma}a_{\mathbf{j}\sigma}
+\left(t^b-\lambda'\bar{u}_\mathbf{i}\right)
b^\dag_{\mathbf{i}\sigma}
b_{\mathbf{j}\sigma}\right]\\
&-&\nonumber \sum_{\mathbf{i}\sigma}\left[\Delta
E+(\lambda_a-\lambda_b
)\bar{u}_\mathbf{i}\right]n_{\mathbf{i},\sigma}^b+H_U\\
&-&\sum_{\mathbf{i},\sigma}(\mu-\lambda_a\bar{u}_\mathbf{i})
\left(n^a_{\mathbf{i}\sigma}+n^b_{\mathbf{i}\sigma}\right)+
\frac{K}{2}\sum_\mathbf{i}\bar{u}^2_\mathbf{i},
\end{eqnarray}
here $\mu$ is the chemical potential.

Hamiltonian Eq.~\eqref{Heff} clearly demonstrates that the  effect
of electron-lattice interaction is two-fold. This interaction in
non-diagonal terms changes the effective bandwidth whereas in
diagonal terms, it shifts the positions of the bands and the
chemical potential. In the earlier analysis~\cite{tb1,bianc}, we have shown that the qualitative features of the phase diagram for the
two-band Hubbard model are mainly determined by two dimensionless
parameters: the ratio of the bandwidths and the relative positions
of the bands. Thus, to construct a minimal model capturing the
main physical effects of electron-lattice interaction, it is
sufficient to keep only $\lambda$ and $\lambda_b$. In addition, we
put $\Delta E=0$ to emphasize the effect of band shift related
only to the electron-lattice interaction. As a result, we get
\begin{equation}\label{ui1} \bar{u}_\mathbf{i}=-\frac{\lambda
z\langle a^\dag_{\mathbf{i}\sigma}a_{\mathbf{j}\sigma}\rangle
+\lambda_b\langle n_\mathbf{i}^b\rangle}{K} \end{equation} and
\begin{eqnarray}\label{Heff1} \nonumber H_{\rm eff}=-\sum_{\langle
\mathbf{i}\mathbf{j}\rangle,\sigma} \left[\left(t^a-\lambda
\bar{u}_\mathbf{i}\right)a^\dag_{\mathbf{i}\sigma}a_{\mathbf{j}\sigma}
+t^b b^\dag_{\mathbf{i}\sigma}b_{\mathbf{j}\sigma}\right]+H_U\\
+ \sum_{\mathbf{i}\sigma}
\lambda_b
\bar{u}_\mathbf{i}n_{\mathbf{i},\sigma}^b
-\sum_{\mathbf{i},\sigma}\mu
\left(n^a_{\mathbf{i}\sigma}+n^b_{\mathbf{i}\sigma}\right)+
\frac{K}{2}\sum_\mathbf{i}\bar{u}^2_\mathbf{i}.\,\,
\end{eqnarray}

Starting from Hamiltonian \eqref{Heff1}, we can point out  the
main qualitative effects of the electron-lattice interaction in
the two-band model. In the absence of doping, $n=0$, bands $a$ and
$b$ are empty and their centers coincide. With the growth of $n$
the wider band $a$ begins to be filled up from the bottom. The band
filling is accompanied by strain $\bar{u}_{\mathbf{i}}=-\lambda
z\langle a^\dag_{\mathbf{i}\sigma}a_{\mathbf{j}\sigma}\rangle/K$.
The average $\langle
a^\dag_{\mathbf{i}\sigma}a_{\mathbf{j}\sigma}\rangle$ is
proportional to the hopping probability and thus is positive. The
strain $\bar{u}_{\mathbf{i}}$ is positive if $\lambda<0$ and
negative if $\lambda>0$. From Eq.~\eqref{Heff1}, it is easy to see
that at any sign of $\lambda$ the bandwidth increases with the
strain. At a certain doping level, the chemical potential attains
the bottom of the narrower band $b$ and this second band starts to
be filled up. In this range of doping, we have two types of the
electrons and the energy of the system depends on $n$ in a rather
complicated manner due to electron-electron correlations. As we
have shown earlier~\cite{tb1}, such situation is favorable for the
phase separation even in the absence of the electron-lattice
interaction.

However, the characteristic feature of the system under study  is
the dependence of the effective band shift $\Delta E_{\rm
eff}=-\lambda_b\bar{u}_{\mathbf{i}}$ on the strain and, hence, on
the doping, according to Eq.~\eqref{ui1}. The sign of the shift
depends on the signs of $\lambda$ and $\lambda_b$. A simple
analysis of Eqs.~\eqref{ui1} and \eqref{Heff1} shows that for the
same signs of $\lambda$ and $\lambda_b$ the value of $\Delta
E_{\rm eff}$ is negative and the sign of $\bar{u}_{\mathbf{i}}$
remains the same at any $n$. If the signs of $\lambda$ and
$\lambda_b$ are different, the strain can change its sign at some
doping level. The change of the sign of the strain results in the
change of the sign of the effective band shift $\Delta E_{\rm
eff}$.

The dependence of $\Delta E_{\rm eff}$ on doping can give rise to
a more sophisticated behavior of the system. If at some doping
level $n^*$, the narrower band crosses the bottom
of the wider band, that is $\lambda_b^2n^*/K\gtrsim zt_a$, then it
could be favorable to have almost all electrons in the $b$ band.
So, there appears a competition between two states with
different values of the strain. It suggests the possibility of a
transition between these two states, which can have a jump-like
form.

In the next section, we present a quantitative analysis of the
possible situations.

\section{Mean field approach}

In the limit of strong electron correlations, $U\rightarrow \infty$, we can describe the evolution of the band structure with the change of $n$ following the method presented in Ref.~\onlinecite{tb1}. We
introduce one-particle Green functions for $a$ and $b$ electrons.
For band $a$, we have
\begin{equation}\label{Grf1}
G_{a\sigma}(\mathbf{i}-\mathbf{j},\,t)=
-i\langle\hat{T}a_{\mathbf{i}\sigma}(t)a^{\dag}_{\mathbf{j}
\sigma}(0)\rangle, \end{equation}
where $\hat{T}$ is the
time-ordering operator. The similar expression can be written for
band $b$. The equation of motion for the one-particle Green function
with Hamiltonian \eqref{Heff1} includes two-particle Green
functions of the form
\begin{equation}\label{tp}
{\cal{G}}_{a\sigma,b\sigma'}(\mathbf{i}-\mathbf{j},\,t)=%
-i\langle\hat{T}a_{\mathbf{i}\sigma}(t)n^b_{\mathbf{i}\sigma'}(t)%
a^{\dag}_{\mathbf{j}\sigma}(0)\rangle\,. \end{equation}
In the considered limit of strong on-site Coulomb repulsion, the presence of two electrons at the same site is unfavorable, and the
two-particle Green function is of the order of $1/U$. The equation
of motion for two-particle Green functions includes the
three-particle terms coming from the commutator of
$a_{\mathbf{i}\sigma}(t)$ or $b_{\mathbf{i}\sigma}(t)$ with the
$U$ terms of Hamiltonian~\eqref{Heff1}, which are of the order of
$1/U^2$ and so on. In these equations, following the Hubbard I
approach~\cite{Hub}, we neglect the terms of the order of $1/U^2$
and use the following decoupling in the Green functions $\langle\hat{T}a_{\mathbf{i}+\mathbf{m}\sigma}(t)n^b_{\mathbf{i}
\sigma'}(t) a^{\dag}_{\mathbf{j}\sigma}(0)\rangle\to\langle
n^b_{\mathbf{i}\sigma'}\rangle\langle\hat{T}
a_{\mathbf{i}+\mathbf{m}\sigma}(t)a^{\dag}_{\mathbf{j}
\sigma}(0)\rangle$. As a result, we derive a closed system for the
one- and two-particle Green functions~\cite{tb1,Hub}. This system
can be solved in a conventional manner by passing from the
time-space $(t,\mathbf{r})$ to the frequency-momentum
$(\omega,\mathbf{k})$ representation. We limit ourselves to
consideration of the case when the total number of electrons
per site does not exceed unity, $n=n^a+n^b\leq1$. The upper
Hubbard sub-bands are empty, and we can proceed to the limit
$U\rightarrow\infty$. In this case, the one-particle Green
functions are independent of $U$ and can be written in the
frequency-momentum representation as~\cite{tb1}
\begin{eqnarray}\label{Ginf} \nonumber
G_{a\sigma}(\mathbf{k},\omega)&=&
\frac{g_{a\sigma}}{\omega+\mu-g_{a\sigma}w_{a}(1-\lambda
\bar{u}/t_a)\zeta(\mathbf{k})}\,,\\
G_{b\sigma}(\mathbf{k},\omega)&=&
\frac{g_{b\sigma}}{\omega+\mu-\lambda_b\bar{u}-g_{b\sigma}
w_{b}\zeta(\mathbf{k})}\,,
\end{eqnarray}
where we put $\bar{u}_\mathbf{i}=\bar{u}$ assuming a homogeneous  strain and introduce the following notation $w_\alpha=zt_\alpha$, $\alpha=a$ or $b$,
\begin{equation}\label{g}
g_{\alpha\sigma}=1-\sum_{\sigma'}n^{\bar{\alpha}}_{\sigma'}
-n^\alpha_{\bar{\sigma}}\,, \end{equation}
$n^\alpha_{\sigma}=\langle n^\alpha_{\mathbf{i}\sigma}\rangle$ is
the average number of electrons per site in state
$(\alpha,\sigma)$, and $\zeta(\mathbf{k})$ is the spectral
function depending on the lattice symmetry. In the considered case
of the simple cubic lattice,  $\zeta({\bf k})=-\left[\cos (k^1d)
+\cos(k^2d)+\cos(k^3d)\right]/3$, $d$ is the lattice constant. In
the main approximation in $1/U$, the magnetic ordering does not
appear and we can assume that
$n^\alpha_{\uparrow}=n^\alpha_{\downarrow}\equiv n^{\alpha}/2$. Below we omit spin indices.

\section{Results and discussion}

Equations \eqref{Ginf} and \eqref{g} demonstrate that the filling
of the bands depends on the strain $\bar{u}$ and the number  of
electrons in one band depends on that in another band. Using the
expression for the density of states $\rho_{\alpha}(E)=-\pi^{-1}
\textmd{Im}\int G_{\alpha}(\mathbf{k},E+i0)d^3{\bf k}/(2\pi)^3$,
we get expressions for the numbers of electrons in bands $a$ and
$b$
\begin{equation}\label{nalpha}
n^{a}\!=\!2g_an_0\!\!\left[\frac{\mu}{g_aw_a\!\left(1\!-
\!\lambda\bar{u}/t_a\right)}\right]\!\!,\,
n^b\!=\!2g_bn_0\!\!\left[\!\frac{\mu\!-\!\lambda_b\bar{u}}{g_bw_b}
\!\right]
\end{equation}
where
\begin{equation}\label{n0}
n_0(\mu')=\int\limits_{-1}^{\mu'}dE'\,\rho_0(E')\,, \end{equation}
and $\rho_0(E')=\displaystyle\int
d^3\mathbf{k}\,\delta(E'-\zeta(\mathbf{k}))/(2\pi)^3$ is the
density of states for free electrons (with the energy normalized
by unity, $|E|\leq1$). The chemical potential $\mu$ in
Eq.~\eqref{nalpha} can be found from the equality
\begin{equation}\label{mun} n=n^a(\mu)+n^b(\mu). \end{equation}
Equations~\eqref{nalpha} include average strain $\bar{u}$, which
itself depends on $n^b$ and an average $\langle
a^\dag_{\mathbf{i}\sigma}a_{\mathbf{j}\sigma}\rangle$. The latter
can be expressed in terms of the Green function as
\begin{eqnarray}\label{avec}
\langle a^\dag_{\mathbf{i}\sigma}a_{\mathbf{j}\sigma}
\rangle&=&-iG_{a\sigma}(\mathbf{i}-\mathbf{j},-0)\\
\nonumber &=&-i\int\frac{d\omega
d^3\mathbf{k}}{(2\pi)^4}G_{a\sigma}(\mathbf{k},
\omega)e^{i\mathbf{k}(\mathbf{i}-\mathbf{j})+i\omega0}. \end{eqnarray}
Thus, we have a system of four
equations~\eqref{ui1}, \eqref{nalpha},  and \eqref{mun} for
finding $\bar{u}$, $n^a$, $n^b$, and $\mu$. These equations are
solved together with the relationships \eqref{Ginf}, \eqref{g},
\eqref{n0}, and \eqref{avec}.

\begin{figure}[t] \begin{center}
\includegraphics[width=0.9\columnwidth]{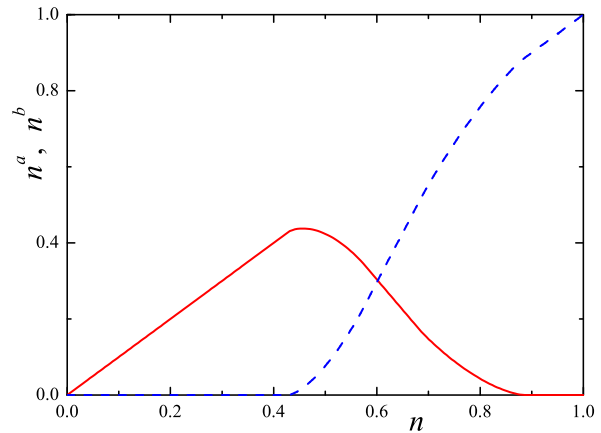} \end{center}
\caption{\label{FigNab} (Color online) Evolution of the occupation
numbers of the $a$ and $b$ bands with doping $n$: $n^a$ (red) solid
line and $n^b$ (blue) dashed line. The parameters are $\lambda/w_a=0.8$,
$\lambda_b/w_a=2.0$, $K=16w_a$ and $w_b/w_a=0.25$.
} \end{figure}
\begin{figure}[t] \begin{center}
\includegraphics[width=0.9\columnwidth]{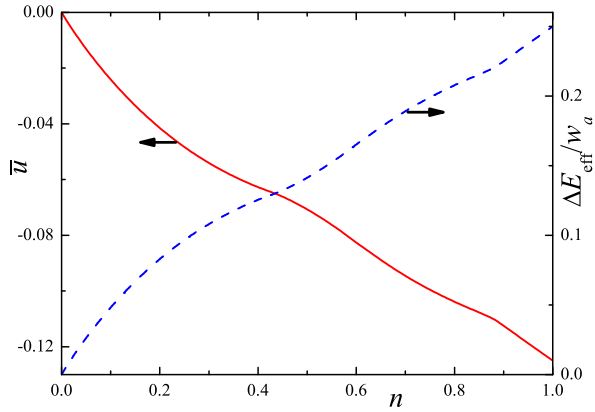} \end{center}
\caption{\label{FigU} (Color online) The dependence of $\bar{u}$
(solid (red) curve) and $\Delta E_{\rm eff}=-\lambda_b\bar{u}$
(dashed (blue) curve) on $n$. The parameters are the same as
in Fig~\ref{FigNab}.} \end{figure}

The dependence of $n^a$ and $n^b$ on doping $n$ is illustrated  in
Fig.~\ref{FigNab}. At low doping, the bottom of $b$ band lies far
above the bottom of the wider band $a$, and there exist only $a$
electrons. The filling of the bands gives rise to a non-zero
strain $\bar{u}$, see Eq.~\eqref{ui1}, and hence to the band shift
$\Delta E_{\rm eff}$. The plots $\bar{u}(n)$ and $\Delta E_{\rm
eff}(n)$ are shown in Fig.~\ref{FigU}. When $\lambda$ and
$\lambda_b$ have the same signs the band $b$ shifts downward
($\Delta E_{\rm eff}>0$). Thus, with the increase of $n$, the
chemical potential crosses the bottom of the $b$ band and $b$
electrons appear in the system. Due to electron-electron
correlations, the effective width of $a$ band starts to decrease.
This band narrowing and increase of $\Delta E_{\rm eff}$ leads to
decreasing of the number of $a$ electrons, and at some doping
level, the charge carriers of type $a$ disappear in the system, see
Fig.~\ref{FigNab}.

\begin{figure}[t] \begin{center}
\includegraphics[width=0.9\columnwidth]{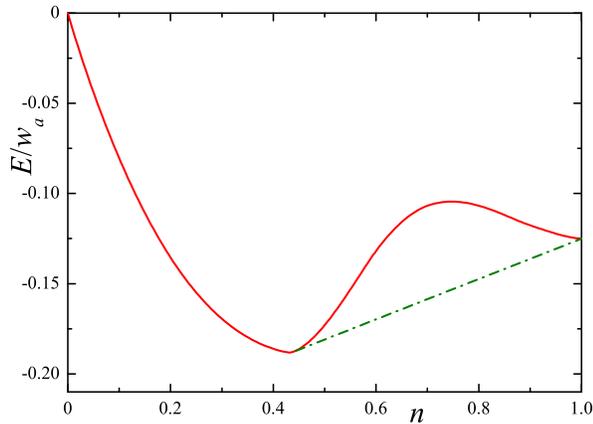} \end{center}
\caption{\label{FigE} (Color online) The energy of the system {\it
vs} doping level $n$. Solid (red) curve corresponds to  the
homogeneous state, whereas the dot-dash (green) curve is the
energy of the phase-separated state. The parameters are the same
as in Fig~\ref{FigNab}.} \end{figure}

The energy of the system in the homogeneous state, \[
E_{\text{hom}}=\sum_\alpha\int \rho_\alpha(E')E'dE', \] is the sum
of electron energies in all filled bands. After straightforward
calculations, we can write $E_{\text{hom}}$ in the form
\begin{eqnarray}\label{Ehom}
\nonumber
E_{\text{hom}}&=&2g_{a}^2w_{a}\left(1-\frac{\lambda \bar{u}}{t_a}\right)
\varepsilon_0\left[\frac{\mu}{g_{a}w_{a}\left(1-\lambda \bar{u}/t_a\right)}\right]\\
&+&2g_{b}^2w_{b}\varepsilon_0\left[\frac{\mu-\lambda_b\bar{u}}{g_{b}w_{b}}\right]
+\lambda_b\bar{u}n_b+\frac{K\bar{u}^2}{2}\,,
\end{eqnarray}
where
\begin{equation}\label{eps}
\varepsilon_0(\mu')=\int\limits_{-1}^{\mu'}dE'E'\,\rho_0(E')\,.
\end{equation}
The dependence of $E_{\text{hom}}(n)$ is shown in Fig.~\ref{FigE} by solid line. We see that within a certain $n$ range the system can have a negative compressibility, $\partial^2E_{\text{hom}}/\partial n^2<0$, which means a possibility for the charge carriers to form two phases with different electron concentrations~\cite{tb1}.

\begin{figure}[t] \begin{center}
\includegraphics[width=0.9\columnwidth]{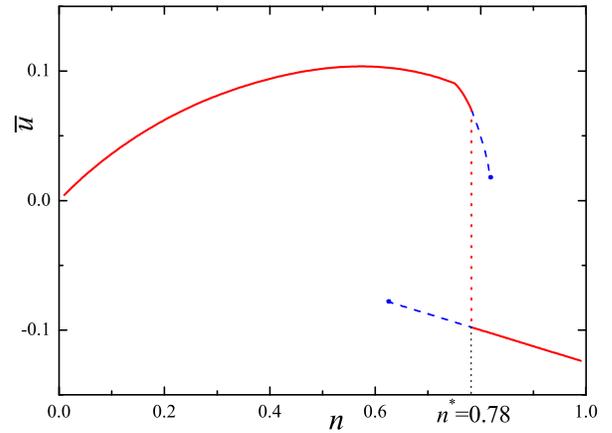} \end{center}
\caption{\label{FigJumpU} (Color online)
Lattice strain as a function of doping. The parameters
are $\lambda/w_a=-1.2$, $\lambda_b/w_a=2$, $K=16w_a$
and $w_b/w_a=0.25$. Jump-like transition between two
states with different values of lattice distortions occurs
at $n=n^*$. Solid (red) lines correspond to the states
with minimum energy, whereas dashed (blue) lines correspond
to metastable states.} \end{figure}
\begin{figure}[t] \begin{center}
\includegraphics[width=0.9\columnwidth]{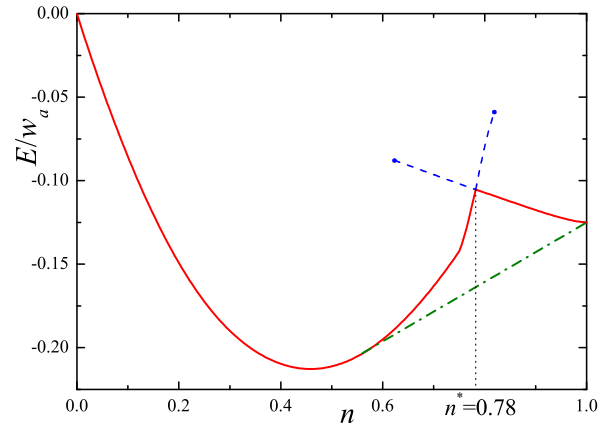} \end{center}
\caption{\label{FigJumpE} (Color online) The energy of the system {\it vs} doping level $n$. Solid (red) curve corresponds to the
homogeneous state, whereas the dot-dash (green)
curve is the energy of the phase-separated state.
The dashed (blue) curves correspond to the energies of
metastable states. The parameters are the same as in
Fig~\ref{FigJumpU}. The kink in the curve $E_{\text{hom}}(n)$
corresponds to the jump-like change in the lattice distortion.}
\end{figure}

The values of parameters chosen to plot Figs.~\ref{FigNab}-\ref{FigE} correspond to a continuous evolution of the strain with doping. However, as it was mentioned in Section II, one could also expect a jump-like transition between states with different values of the strains at certain values of parameters. Such a situation is illustrated in Figs.~\ref{FigJumpU} and \ref{FigJumpE}. In Fig.~\ref{FigJumpU}, we can see that at some $n=n^*$ the strain $\bar{u}$ exhibits a stepwise transition with the change of the sign. In the vicinity of $n^*$, there exist two competing states, $A$ and $B$, corresponding to two solutions of the system of
equations~\eqref{ui1}, \eqref{nalpha}, and \eqref{mun} for
$\bar{u}$, $n^a$, $n^b$, and $\mu$. In the state $A$, we have
$n^b\ll n^a$ or $n_b=0$, whereas in the $B$ state, $b$ electrons are
prevailing. The energies of these states coincide at $n=n^*$ and
at $n>n^*$, state $A$ has the a higher energy than state $B$. The
minimum energy of the homogeneous state is shown in
Fig.~\ref{FigJumpE} by solid line and the energies of the
metastable states are depicted by dashed lines. The change in the
type of state corresponds to the kink in $E_{\text{hom}}(n)$ curve
at $n=n^*$.

\begin{figure}\centering
\subfigure{\includegraphics[width=0.5\textwidth]{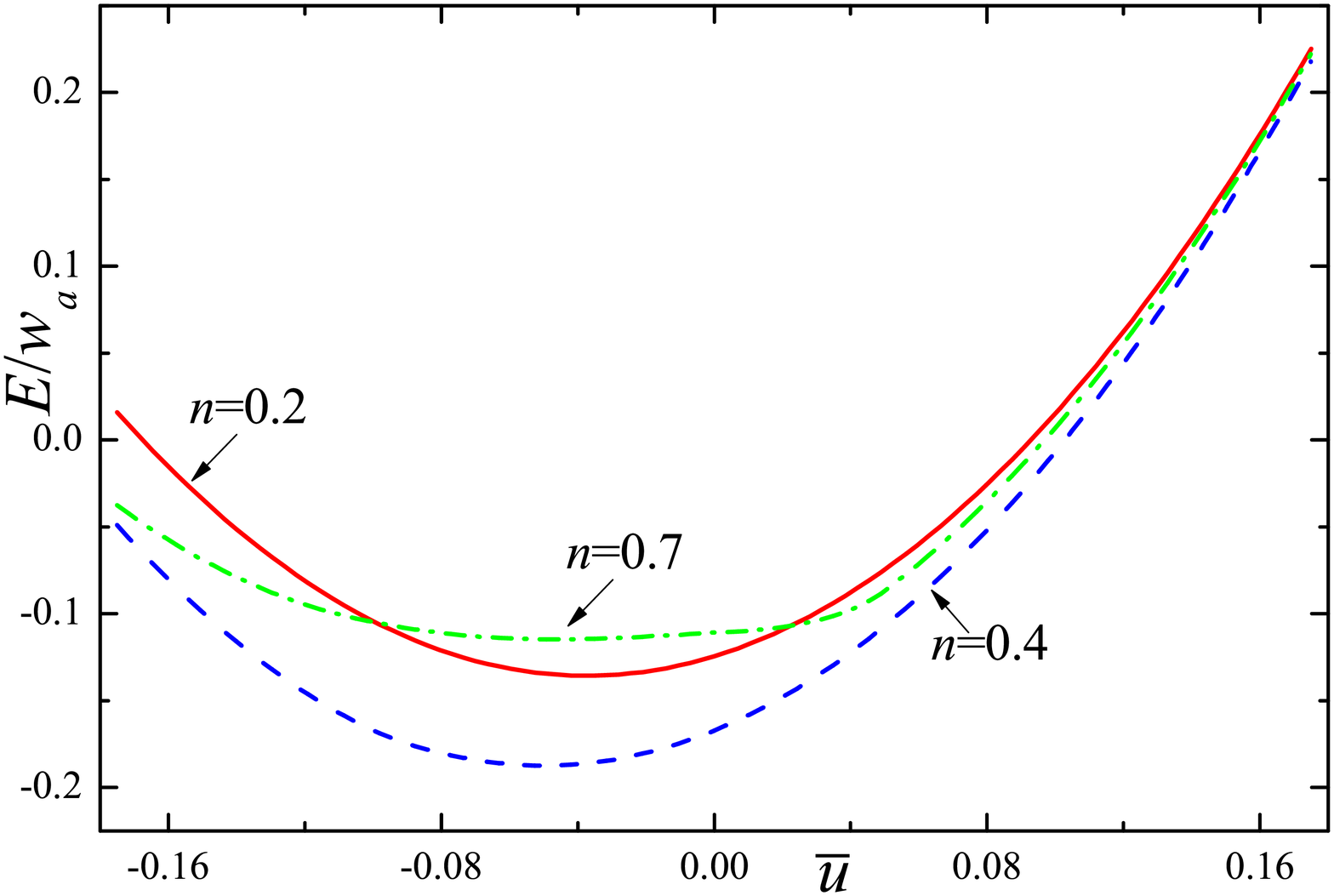}}\\
\subfigure{\includegraphics[width=0.5\textwidth]{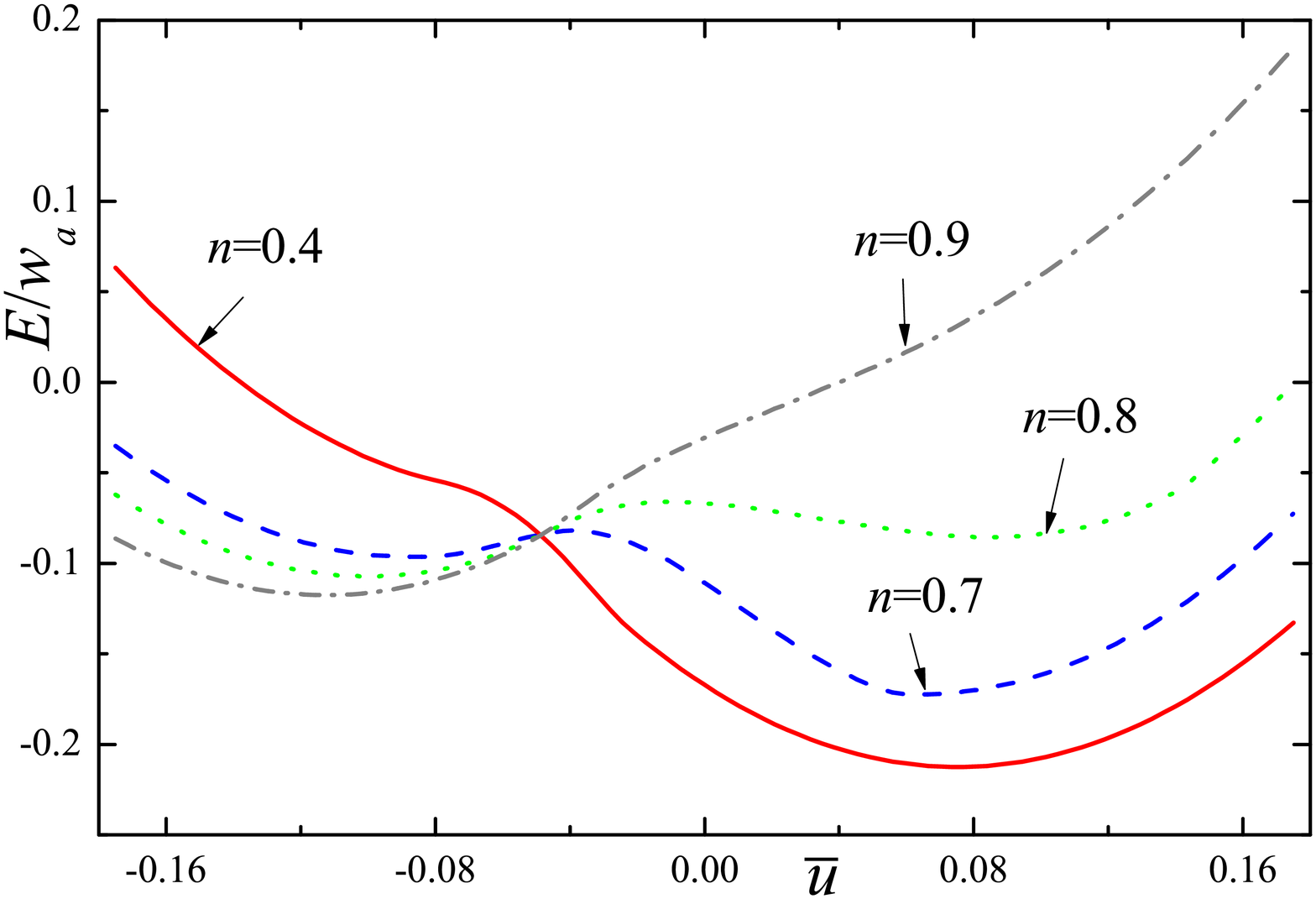}}
\caption{The energies of the homogeneous state depending
on the strain $u$ calculated for $\lambda/w_a=0.8$ (upper panel),
and for $\lambda/w_a=-1.2$ (lower panel) at different
values of doping $n$. Other parameters are the same for both
panels; $\lambda_b/w_a=2$, $K=16w_a$, and $w_b/w_a=0.25$.
In the case of $\lambda/w_a=-1.2$, the function $E_{hom}(u)$
has two minima within a certain doping range.}\label{FigEvsU}
\end{figure}

The existence of two competing states in some range of
parameters can be illustrated in the following way. Let us
study Hamiltonian~\eqref{Heff1} where the strain $\bar{u}=u$ is
considered as an independent parameter. We analyze this Hamiltonian
in the way similar to that described above. Namely, at each given $u$, we solve the system of equations \eqref{nalpha} and \eqref{mun}
for $n^a$, $n^b$, and $\mu$, and then find the system energy per
lattice site $E_{hom}(u)$. The optimum value $\bar{u}$ is then determined by minimization of $E_{hom}(u)$. The numerical analysis
shows that the function $E_{hom}(u)$ has one or two minima
depending on model parameters. The functions $E_{hom}(u)$
calculated for two sets of parameters $\lambda$, $\lambda_b$, $K$
and $w_b$ at different doping levels $n$ are shown in
Figs~\ref{FigEvsU}. At $u=\bar{u}$ corresponding to the minimum of
$E_{hom}(u)$, we have
\begin{equation} \frac{\partial
E_{hom}}{\partial u}\equiv\frac{1}{N}\left\langle\frac{\partial
H_{\text{eff}}}{\partial u}\right\rangle= \lambda z\langle
a^\dag_{\mathbf{i}\sigma}a_{\mathbf{j}\sigma}\rangle+
\lambda_bn^b+K\bar{u}=0\,, \end{equation}
and we come back to
Eq.~\eqref{ui1} for $\bar{u}$.

\begin{figure}[t]\centering
\subfigure{\includegraphics[width=0.5\textwidth]{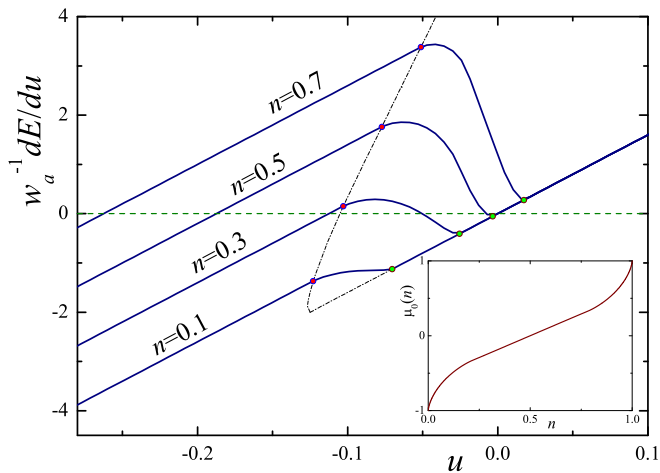}}
\caption{The dependence of $E'(u)$ on the strain $u$
at different values of doping $n$. The parameters are
$\lambda=0$, $\lambda_b/w_a=6$, $K=16w_a$, and $w_b/w_a=0.25$.
The values of $u_1(n)$ and $u_2(n)>u_1(n)$ are marked by circles.
In the inset, the function $\mu_0(n)$, inverse to the
function~\eqref{n0} is shown.}\label{FigDEvsU}
\end{figure}

In the general case, it is difficult to find explicit conditions for the existence of the jump-like transition. Here, we analyze the
important particular case of $\lambda=0$. Let us consider the
function $E'(u)\equiv\partial E_{hom}/\partial u$, which now has a
form $E'(u)=Ku+\lambda_bn^b(u)$. The $E'(u)$ curves calculated at
different values of doping are illustrated in Fig.~\ref{FigDEvsU}.
For large negative strain, the $b$ band lies far below the $a$
band, and we have $n^a=0$ and $n^b=n$. Thus, $E'(u)=Ku+\lambda_bn$
linearly grows with $u$ up to the point $u=u_1$, when the
chemical potential $\mu$ reaches the bottom of the effective $a$
band $-g_awa=-(1-n)w_a$, and $a$ electrons appear in the system.
Using second equation in the system Eq.~\eqref{nalpha} with
$n_b=n$, and $\mu=-(1-n)w_a$, we obtain the following expression
for $u_1$:
\begin{equation}\label{u1}
u_1=\frac{-1}{\lambda_b}\left[(1-n)w_a+\left(1-\frac{n}{2}\right)
w_b\mu_0\left(\frac{n}{2-n}\right)\right],
\end{equation} where $-1<\mu_0(n)<1$ is the function inverse to
$n_0(\mu')$, Eq.~\eqref{n0}. This function is shown in the inset
in Fig.~\ref{FigDEvsU}. For large positive $u$, when band $b$
lies above band $a$, we have $n^a=n$ and $n^b=0$, and
consequently $E'(u)=Ku$ linearly decreases with $u$ till the point $u_2>u_1$, where $b$ electrons appear. Acting in a
similar way, we find
\begin{equation}\label{u2}
u_2=\frac{1}{\lambda_b}\left[(1-n)w_b+\left(1-\frac{n}{2}\right)
w_a\mu_0\left(\frac{n}{2-n}\right)\right].
\end{equation}

It is clearly seen from Fig.~\ref{FigDEvsU}, that if
\begin{equation}\label{jucond0} E'(u_1)=Ku_1+\lambda_b
n\geq0,\;\;E'(u_2)=Ku_2\leq0\,, \end{equation}
then the function
$E'(u)$ has three zeros, and the energy $E(u)$, in turn, has two
minima. Using Eqs.~\eqref{u1} and~\eqref{u2}, and taking the
equality signs in relations~\eqref{jucond0}, we get the estimate
for the minimum value of $\lambda_b$, at which the jump-like
transition can occur
\begin{equation}\label{jucond}
\frac{(\lambda_b^*)^2}{K}=\frac{1-n_2}{n_2}w_a\left(1-\frac{w_b^2}{w_a^2}\right), \end{equation}
where $n_2$ is
found from the equation
\begin{equation}
\mu_0\left(\frac{n_2}{2-n_2}\right)=-\frac{1-n_2}{1-n_2/2}\frac{w_b}{w_a}.
\end{equation}
The value of $\lambda_b^*$ decreases with $w_b$. For very small ratio $w_b/w_a\ll1$, we have
$\mu_0[n_2/(2-n_2)]\approx0$, that is, $n_2\approx2/3$, and
\begin{equation}\label{lambdac}
\lambda_b^{*}\approx\sqrt{\frac{w_aK}{2}}. \end{equation}
Note that the function $E(u)$ can still have two minima when
conditions~\eqref{jucond0} are not met, since the derivative
$E'(u)$ can continue to grow (decrease) above (below) $u_1$
($u_2$), and, consequently, $\lambda_b^{*}$ found from these
conditions overestimates its value.

For $\lambda=0$ the jump-like transition occurs for relatively
large values of $\lambda_b$. For example, at $K=16w_a$ and
$w_b=0.25w_a$, from Eq.~\eqref{jucond} one obtains
$\lambda_b^{*}\approx3.6w_a$. The numerical analysis shows,
however, that even small negative (if $\lambda_b>0$) $\lambda$
sufficiently reduces the threshold value of $\lambda_b^*$. For
example, at $\lambda=-0.4w_a$, the jump-like behavior arises
starting from $\lambda_b^{*}\approx2.0w_a$ ($K=16w_a$ and
$w_b=0.25w_a$). Thus, different signs of $\lambda$ and
$\lambda_b$ favor the existence of such transition.

\section{Conclusions}

The electron-lattice interaction plays an important role in the systems with strongly correlated electrons affecting their behavior with doping. The interaction of electrons with the lattice distortions results, first, in the hopping probability and, second, in the relative shifts of the electronic bands. We analyzed the problem in the framework of the two-band Hubbard model. We demonstrated that if the electron-lattice interaction is strong enough, there appears a competition between states with different values of strains and the transition between these states can occur in a jump-like manner. We also showed that the electron-lattice interaction produces a pronounced effect on the conditions of the electronic phase separation since it influences the value of the bandwidth ratio and the relative positions of the bands.

\section*{Acknowledgements}

The work was supported by the Russian Foundation for Basic
Research (projects 07-02-91567 and 08-02-00212).

\end{document}